\documentclass[aps,prb,amsmath,twocolumn,amssymb,titlepage]{revtex4-2}
\usepackage[T1]{fontenc}
\usepackage{mathrsfs}
\usepackage{amssymb}
\usepackage{amsmath}
\usepackage{graphicx}
\usepackage{braket}
\usepackage{physics}
\usepackage{bm}
\usepackage{bbold}
\usepackage{notes2bib}
\usepackage[utf8]{inputenc}
\usepackage{color}
\usepackage[colorlinks=true,citecolor=blue,urlcolor=blue]{hyperref}

\begin{document}
	
	\title{The emergence of interface states in graphene/transition metal dichalcogenides heterostructure with lateral interface}
	\author{Zahra Khatibi}
	\affiliation{School of Physics, Trinity College Dublin, Dublin 2, Ireland}
	\author{Stephen R. Power}
	\email{stephen.r.power@dcu.ie}
	\affiliation{School of Physical Sciences, Dublin City University, Ireland}
	\affiliation{School of Physics, Trinity College Dublin, Dublin 2, Ireland}
	
	\date{\today}
	
	\begin{abstract}
     The relative strength of different proximity spin-orbit couplings in graphene on transition metal dichalcogenides (TMDC) can be tuned via the metal composition in the TMDC layer. 
     While Gr/MoSe$_2$, has a normal gap, proximity to WSe$_2$ instead leads to valley-Zeeman-driven inverted bands. Although the $\mathbb{Z}_2$ index vanishes, these systems enable a concentration-dependent topological crossover with band gap closure when graphene is stacked on a composite or alloyed TMDC layer.
     This is due to a nonzero Berry curvature at the individual valleys and a change of the valley Chern index at a critical composition ratio. 
     Therefore, inherently, we also expect that stacked heterostructures of graphene on composite TMDC layers should host localised boundary modes due to the presence of Gr/WSe$_2$- and Gr/MoSe$_2$-like domains with opposite valley Chern indices. 
     In this study, we show that a Gr/(Mo--W)Se$_2$ heterostructure with a lateral interface in the TMDC layer can indeed host topologically protected in-gap propagating modes, similar to those at the border of commensurate AB and BA domains in biased minimally-twisted bilayer graphene.
     However, the stability of these modes depends crucially on the system size. 
     We demonstrate that the electronic behaviour of Gr/(Mo--W)Se$_2$ heterostructures evolves from a homogeneous effective medium to a superposition of domain-localised bands and zero-energy branch crossings as the domain size in the alloyed TMDC layer is increased. 
	\end{abstract}
	\maketitle
	
	\section{Introduction}
    
    Graphene-based systems with broken inversion symmetry are host to long sought topologically-protected helical (TPH) states \cite{zarenia2011chiral,qiao2011electronic,klinovaja2012helical,shevtsov2012graphene,ju2015topological,li2016gate,yin2016direct,tsim2020perfect,kwan2021domain}. 
    For instance, twisted bilayer graphene has recently seen renewed interest for hosting the quantum valley Hall effect and the realization of one-dimensional topological modes. 
    These transport channels emerge in long-ranged moir\'e structures of twisted bilayer graphene where a network of domain walls form in between commensurate AB and BA stacking regions \cite{san2013helical,huang2018topologically,rickhaus2018transport,yoo2019atomic,lebedeva2019energetics}. An external electric field lifts the inversion symmetry, giving rise to a nonzero Berry curvature within, and TPH states along, the spatially varying stacking registries \cite{efimkin2018helical}.
    Similar localised subgap states have been shown to emerge in a uniform bilayer, and later in chirally stacked N-layer, with opposite polarity-gate domains \cite{martin2008topological,yao2009edge}.
    In fact, the low-energy effective Hamiltonian of multi-layer graphene under a uniform vertical electric field has a finite Chern number ($1/2$) per spin with an opposite sign at each valley \cite{jung2011valley}.
    The valley-projected Chern index changes sign over a boundary between domains with opposite polarity, imposing a band gap closure and the emergence of valley helical edge states. 
    Overall, dictated by the valley Chern index (i.e. $C_V=C_K-C_{K^\prime}$, the difference between the valley-projected Chern indices), the local boundary states consist of  
    N branches of interface states in chirally stacked N-layer graphene with opposite polarity-gate domains.
    
    Exotic states with built-in spin and valley properties are also predicted to emerge in graphene when spin-orbit interactions are included. With an enhanced homogeneous Kane-Mele, or ìntrinsic', spin-orbit coupling (SOC), a quantum spin Hall edge state emerges near charge neutrality \cite{kane2005quantum}. This topologically-protected edge state consists of counter-propagating channels with opposite spins \cite{kane2005z}.  
    Proximity effects induced in graphene by transition metal dichalcogenides (TMDCs) can also generate different quantum Hall effects \cite{avsar2014spin,han2014graphene,wang2015strong,wang2016origin,garcia2017spin,ghiasi2017large,dankert2017electrical,volkl2017magnetotransport,friedman2018spin,wakamura2018strong,benitez2018strongly,offidani2018microscopic,island2019spin,david2019induced,arora2020superconductivity}. 
    These effects are rather complex and involve an interplay between a diverse range of sublattice-dependent intrinsic, or `valley Zeeman' (VZ), and Rashba-type SOC \cite{min2006intrinsic,Gmitra2016trivial,yang2016tunable,yan2016two,Tobias2017Magnetotransport,sierra2021van,Tiwari2021Electric}. 
    The Rashba term couples electrons of different spin orientations introducing in-plane spin precession effects in the absence of external magnetic fields \cite{cummings2017giant,yang2017strong,Garcia2018Spin,Josep2021Electrical}.
    Although the staggered nature of the VZ term in proximitised graphene leads to a trivial topological order \cite{Alsharari2018Topological,Frank2018Protected}, the nonzero Berry curvature at individual valleys enables a topological interpretation of the electronic structure in graphene on TMDCs \cite{Alsharari2016mass}. 
    For instance, the band-inverted regime in Gr/WSe$_2$, where the VZ term dominates, enables the formation of topologically-protected pseudohelical modes in finite-sized ribbons \cite{Frank2018Protected}. 
    Recent studies also suggest the emergence of pair of interface states per valleys in proximitised graphene where the VZ term changes sign \cite{touchais2022robust}. 
    The interplay of a changing sign VZ and constant Rashba SOC in graphene can be mapped to stacking registry domains in bilayer graphene where topological channels were previously shown to exist \cite{vaezi2013topological,zhang2013valley}.    
    
    Previously we have shown that proximity spin-orbit coupling in graphene on TMDCs can be controlled via the metal composition in the TMDC layer \cite{khatibi2022proximity}. 
    In a composite heterostructure, the electronic dispersion resembles a uniform Gr/TMDC system with a weighted average of the SOC parameters of the pure systems, i.e. the weighted average (WAVG) model bands. 
    Particularly, in the case of composite ${\rm Gr/W_{\chi}Mo_{1-\chi}Se_2}$ system, where ${\chi}$ denotes the tungsten-molybdenum ratio, the change of the valley Chern index over the critical composition ratio $\chi = 0.33$ leads to band gap closure and a topological transition.
    This is because the individual Gr/MoSe$_2$, and Gr/WSe$_2$ systems host direct and inverted band regimes respectively due to the relative strengths of the mass and VZ terms (See Fig. \ref{schematic}(b,d)). Therefore, a crossover in the magnitudes of the two terms can lead to a topological crossover in the composite system \cite{khatibi2022proximity}. 
    We should also therefore expect composite Gr/TMDC heterostructures to host localised boundary states where domains with opposite valley Chern indices are formed. 
    This corresponds, in our case, to a Gr/(Mo--W)Se$_2$ heterostructure with a lateral interface in the TMDC layer.
    The crossover between the two unequal valley Chern indices occurs at the border of the Gr/MoSe$_2$ and Gr/WSe$_2$ domains where the direct and inverted band regimes meet.    
    
    In this study, using the tight-binding (TB) model, we investigate the emergence of TPH states in Gr/TMDC heterostructures 
    at the border of the broad MoSe$_2$ and WSe$_2$. 
    We observe a similar behaviour to the case of bilayer graphene with different stacking registries or polarity-gate domains where the topological branch crossing leads to band gap closure and the emergence of zero-energy modes.
    Previous studies show that the boundary modes only emerge in twisted bilayer graphene with remarkably small twist angles where large wavelength moir\'e pattern form \cite{san2013helical,huang2018topologically,efimkin2018helical}. 
    Similarly, here we show that the presence of the interface modes is closely tied to the Gr/MoSe$_2$ and Gr/WSe$_2$ domain size. 
    We find that in order for zero-energy branch crossings to emerge, the domain size should be comparable to Fermi wavelength, $\lambda_F \approx hv_F/E$, where $E$ is the energy scale of the system.
    Our findings are experimentally viable by recent advances in the large-scale fabrication and characterization of lateral TMDC heterostructures via controlled alloying \cite{sahoo2018one,liu2018interface,zheng2018band,sahoo2019bilayer,wang2021stacking,nugera2022bandgap} and chalcogen monomer feeding techniques \cite{zuo2022robust}. Consequently, this suggests the deliberate alloying and patterning of the TMDC layer as an effective mechanism for tuning the topological behaviour of graphene and the realization of topologically protected transport channels for spintronic applications.
    
    This paper is organized as follows: In section \ref{tb_model}, we discuss the details of our TB model and the geometry of the system.
    In section \ref{DFT_WAVG}, we present the density functional theory (DFT) calculated electronic structure and look for signatures of interface states in the band dispersion. Next, we investigate the size effects on the localised interface states in section \ref{size_eff} and find the critical system size for a realistic Gr/(Mo--W)Se$_2$ heterostructure that supports the formation of well-defined boundary modes.
    In section \ref{DW_modes}, we focus on the detailed properties of interface modes and their evolution with system size. 
    Finally, we conclude our findings in section \ref{conclusions}.

	\section{model}\label{tb_model}
	To study the interface states we start with  the TB Hamiltonian of $p_z$ orbitals in a graphene layer on a TMDC layer \cite{Alsharari2016mass,Kochan2014spin,Gmitra2016trivial}:
	\begin{eqnarray}\label{Eq:TB}
		\mathcal{H} &=&
		\sum_{\langle i,j\rangle,\sigma} t~ c_{i\sigma}^\dagger c^{\phantom\dagger}_{j\sigma}+
		\sum_{i,\sigma} ~\Delta~ \xi_{c_i}\,c_{i\sigma}^\dagger c^{\phantom\dagger}_{i\sigma} \\
		&&+\frac{2i}{3}\sum_{\langle i,j\rangle}\sum_{\sigma,\sigma'}c_{i\sigma}^\dagger c^{\phantom\dagger}_{j\sigma'}\left[\lambda_{\rm R}^{ij} \left(\mathbf{\hat{s}}\times \mathbf{d}_{ij}\right)_z\right]_{\sigma\sigma'}\nonumber \\  
		&&+\frac{i}{3}\sum_{\langle\langle i,j\rangle\rangle}\sum_{\sigma,\sigma'}c_{i\sigma}^\dagger c^{\phantom\dagger}_{j\sigma'} \left[\frac{\tilde\lambda_{\rm I}^{ij}}{\sqrt{3}}\nu_{ij}\hat{s}_z \right]_{\sigma\sigma'}.\nonumber
	\end{eqnarray}

	\begin{table}[bpt]
		\begin{center}
			\begin{tabular}{cccccc}
				\hline\hline
				TMDC &  $t$ & $\Delta$ & $\lambda_{\rm R}$  & $\lambda_{\rm I}$ & $\lambda_{\rm VZ}$  \\
				&[eV] & [meV] &  [meV] & [$\mu$eV] & [meV] \\ \hline
				
				\hline
				MoSe$_2$	&	2.53	&	-0.59	&	0.29	&	-3.87	&	0.28\\
				WSe$_2$	&	2.531	&	-0.52	&	0.51	&	-3.06	&	1.15\\ 
				\hline\hline
			\end{tabular}
		\end{center}
		\caption{{\small Calculated orbital and spin-orbit parameters of graphene in a Gr/TMDC heterostructures, found by fitting DFT results to a continuum Dirac model. $t$ is the nearest neighbour tunneling energy, $\Delta$ is the proximity-induced orbital gap, $\lambda_{\rm I}$ is the intrinsic spin-orbit coupling, $\lambda_{VZ}$ is the valley Zeeman spin-orbit coupling, and $\lambda_{\rm R}$ is the Rashba SOC.}} \label{Tab:param}
	\end{table}

	Here, the $c_{i\sigma}^\dagger(c_{i\sigma})$ operator creates (annihilates) an electron at atomic site $i$ with spin $\sigma$.
	$\xi_{c_i} = \pm 1 $ is a sublattice index and $\hat{s}$ is the spin vector made of  Pauli matrices.
	$\mathbf{d}_{ij}$ and $\mathbf{D}_{ij}$ are the unit vectors connecting the nearest neighbour and next-nearest neighbours, respectively, and $\nu_{ij}=1(-1)$ defines the trajectory sign, i.e. clockwise (counterclockwise) from  the site $j$ to  site $i$. $t$ is the spin-independent hopping and $\Delta$ denotes the mass term.
    $\lambda_{\rm R}$ is the Rashba coupling, i.e. a substrate-induced SOC due to symmetry-breaking in the transverse direction that introduces in-plane spin textures.
    $\tilde \lambda_{\rm I}$ is the generic sublattice-dependent intrinsic parameter that can be written in terms of uniform parameters of Dirac Hamiltonian \cite{khatibi2022proximity}:
	\begin{align*}
		\tilde \lambda_{\rm I}^A &= (\lambda_{\rm I}+\lambda_{\rm VZ}) \\
		\tilde\lambda_{\rm I}^B &= (\lambda_{\rm I}-\lambda_{\rm VZ})
	\end{align*}
    where the $\lambda_{\rm I}$ and $\lambda_{\rm VZ}$ are the intrinsic and VZ terms that can be thought of as the sublattice symmetric and asymmetric contributions to a Kane-Mele type coupling.
	Note that there are additional terms in proximitised graphene Hamiltonian, namely, $\mathcal{H}_{\rm PIA}$ and $\mathcal{H}_{\Delta {\rm PIA}}$, that are responsible for the renormalisation of the Fermi velocity and spin-dependent band splitting further from the valleys \cite{Kochan2014spin}. 
 However, since their contribution to the low-energy electronic behaviour close to valleys is negligible compared to other SOC terms, they are not included in this study.

    \begin{figure}[tp]
    	\center
        \includegraphics[width=.85\linewidth]{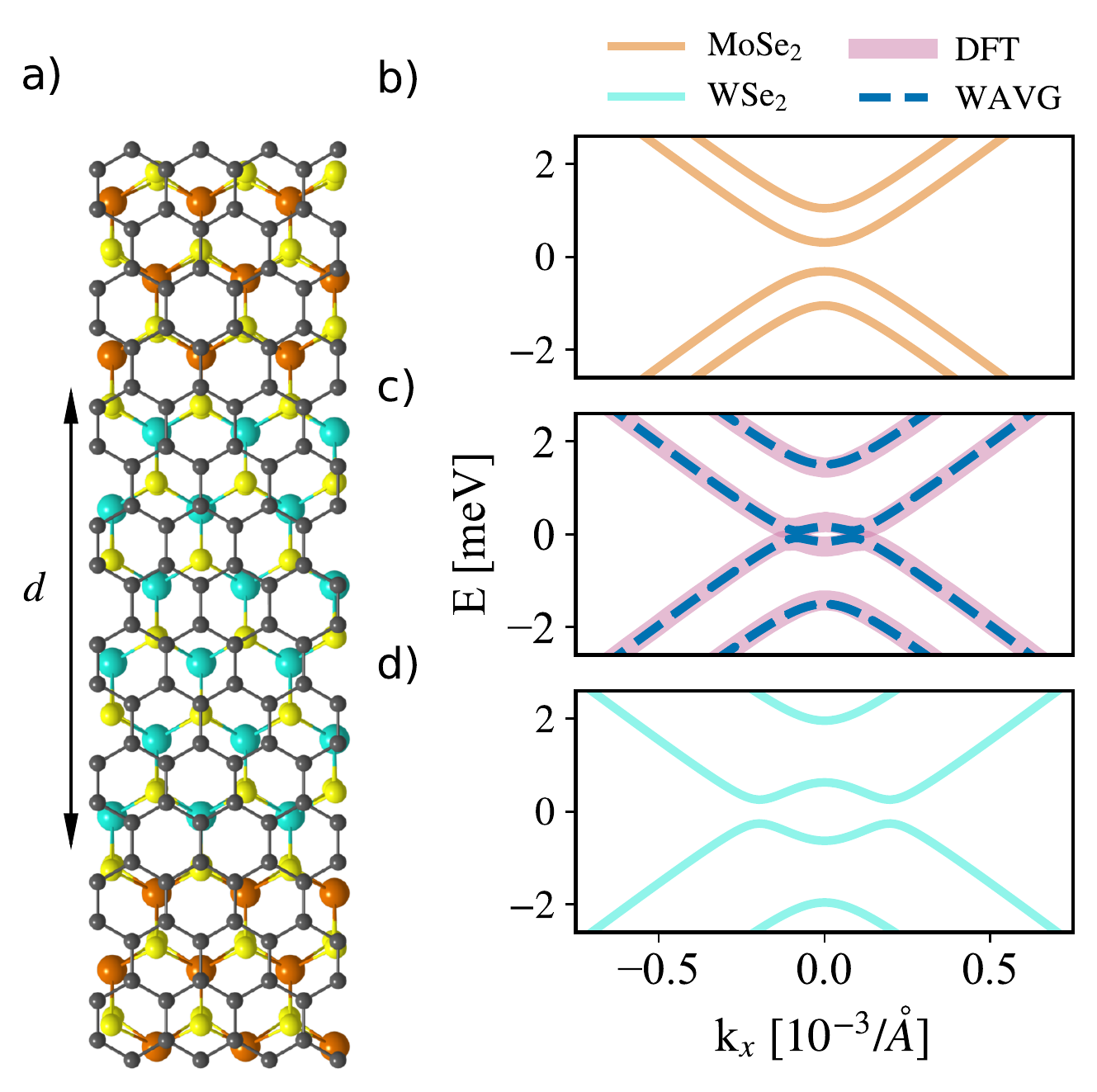}
    	\caption{\small (a) DFT cell of a Gr/(Mo--W)Se$_2$ heterostructure with lateral interfaces in the TMDC layer. Tungsten, molybdenum, and carbon atoms are indicated by cyan, orange, and black spheres. The cell is equally divided into two Gr/MoSe$_2$ and Gr/WSe$_2$ regions. The interfaces between the W- and Mo- like regions are aligned with the zigzag direction. Electronic dispersion of a pristine (b)  Gr/MoSe$_2$, (d) Gr/WSe$_2$ heterostructure. (c) DFT (solid), and homogeneous WAVG model bands (dashed) energy dispersion of a Gr/(Mo--W)Se$_2$ heterostructure showed in (a) near the K valley.
    	\label{schematic}} 
    \end{figure}

    To model the interface modes in proximitised graphene, we consider a TMDC layer that is composed of MoSe$_2$ and WSe$_2$ domains, which should guarantee a crossover from normal to inverted bands and lead to the emergence of TPH states due to the switching sign of the valley-Chern index. 
    To allow for periodic boundary conditions, we consider a Gr/(Mo--W)Se$_2$ supercell with two interfaces between the Gr/WSe$_2$ and Gr/MoSe$_2$ regions as shown in Fig.\ref{schematic}(a). 
    Note that the exact system illustrated here is only used for DFT calculations, while for the TB simulations, we consider significantly larger, but schematically similar, supercells.
    We use the SOC parameters of Gr/MoSe$_2$ (Gr/WSe$_2$) detailed in Tab. \ref{Tab:param} for carbon atoms located in the MoSe$_2$ (WSe$_2$) domain. 
    Note that we do not include the sublattice-symmetric intrinsic SOC as it is significantly smaller than the other SOC parameters.
    Furthermore, we consider the interfaces aligned with the zigzag direction of the graphene lattice.
    As a result, the Dirac points are projected onto two distinct momenta, which prevents intervalley scattering that is detrimental to the interface modes \cite{touchais2022robust}. 
    Note that we present results for abrupt step-like transitions between the proximity SOC of the neighbouring domains, however, calculations with a smoothly-changing SOC over the interfaces yield similar results.

    \section{Alloyed and interface states}\label{DFT_WAVG}

    Before considering large-scale TB simulations, we first consider a more atomically accurate DFT computation of a Gr/(Mo--W)Se$_2$ heterostructure using the \textsc{Quantum} ESPRESSO package \cite{Giannozzi_2009}.  
    We consider a supercell of 3.4 nm width with commensurate 4:3 Gr/TMDC ratio structure. The supercell is equally divided into Gr/WSe$_2$ and Gr/MoSe$_2$ domains as shown in Fig. \ref{schematic}(a).
    The details of DFT calculations can be found in the footnote \cite{DFT}.
    \begin{figure*}[tp]
    	\center
    	\includegraphics[width=\linewidth]{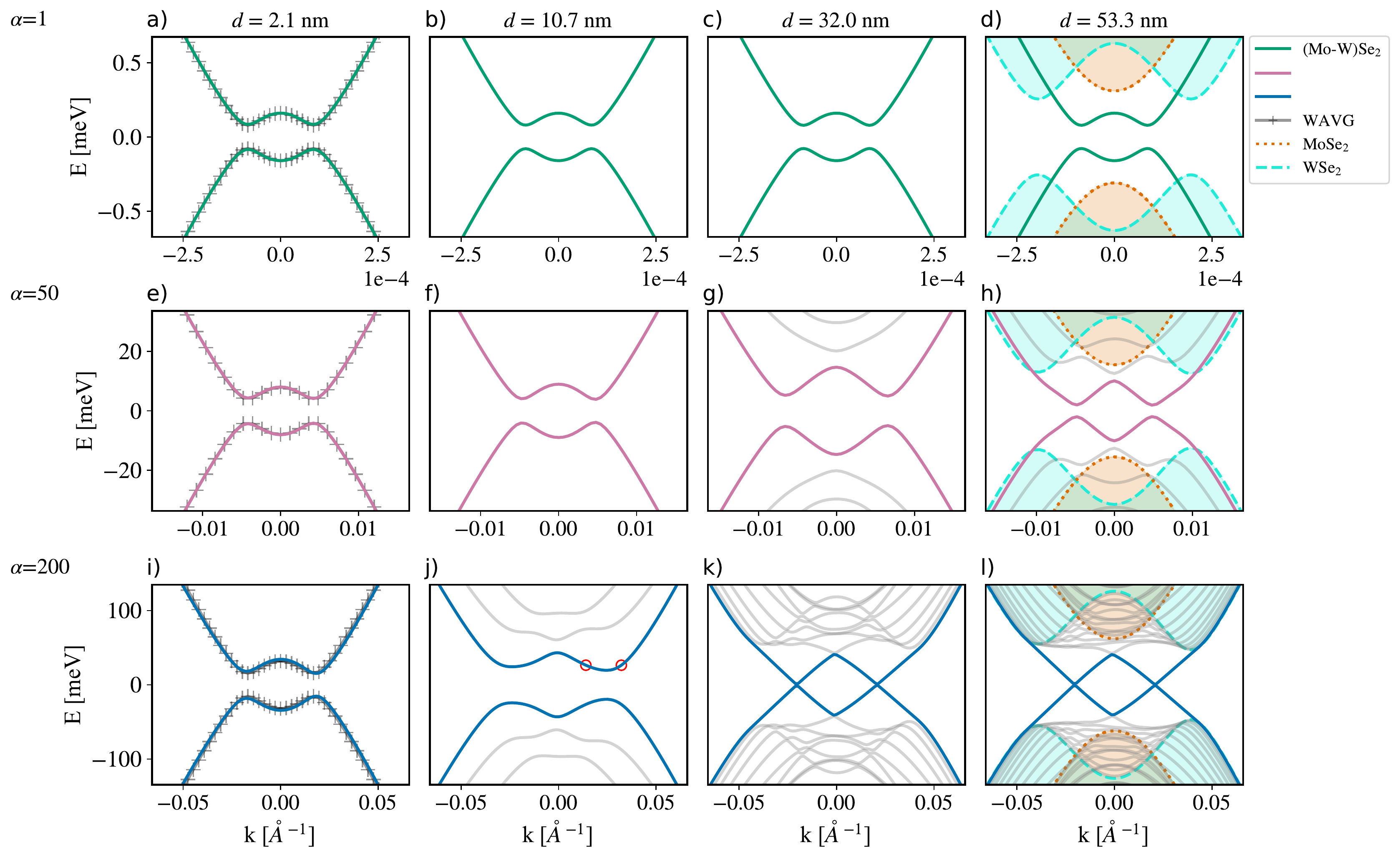}
    	\caption{\small The evolution of electronic band structure of a proximitised graphene with different SOC domains versus the SOC parameter and domain size. The outermost conduction and valence bands near the K valley are illustrated in  bold green, purple, and blue to highlight the emergence of the interface states. The SOC parameters used are those of Gr/MoSe$_2$, and Gr/WSe$_2$ enhanced by a factor of (a--d) $\alpha=1$ (not enhanced), (e--h) $\alpha=50$, and (i--l) $\alpha=200$. The WAVG model bands and those of a pristine Gr/TMDC heterostructure are depicted in the first and last columns for comparison. The lattice size is indicated at top of each column. The empty red disks in (j) are the specific states that will be used in Fig. \ref{wf} for real space expansion of the wave functions. 
    	\label{band_transform}} 
    \end{figure*}
    The DFT calculated electronic dispersion of the Gr/(Mo--W)Se$_2$ heterostructure is illustrated in Fig.\ref{schematic}(c). 
    The resulting band structure does not correspond directly to either of the pure Gr/MoSe$_2$ (Fig.\ref{schematic}(b)) or Gr/WSe$_2$ (Fig.\ref{schematic}(d)) heterostructures, and furthermore, does not show the band gap closure or branch crossings indicative of localised interface modes. 
    Instead, it closely resembles the band structure predicted by the effective WAVG model (dashed line), i.e. a Dirac model using the weighted average of the SOC parameters from each of the pure systems (See Tab.\ref{Tab:param}) \cite{khatibi2022proximity}. 
    The electronic dispersion shows an inverted band regime, as expected for an alloyed ${\rm Gr/W_{\chi}Mo_{1-\chi}Se_2}$ system with $\chi=0.5$ \cite{khatibi2022proximity}. 
    Although we have distinct Mo and W domains in the system, the band dispersion is identical to that of an alloyed system with the same concentration of randomly-distributed metal atoms.
    This also suggests a bulk band gap closure should be expected when the Mo region expands to $66\%$ of the TMDC layer, as this is the critical concentration ratio in random alloys. 
    The similarity between the DFT and effective WAVG model, and the lack of interface states, can be attributed to a strong overlap between the wave functions of the Gr/MoSe$_2$ and Gr/WSe$_2$ domains. This prevents states with different valley Chern indices from forming in each domain, and suppresses interface states along the domain boundaries.    
    However, we note here that the domains are relatively small compared to the Fermi wavelength dictated by the relevant energy scale, namely the proximity-induced energy gap, which from the continuum model Hamiltonian is $\sim 2|\Delta+\lambda_{\rm VZ}|$ \cite{Gmitra2016trivial} where $\Delta$ and $\lambda_{\rm VZ}$ are the magnitudes of staggered onsite potential and VZ SOC parameter, respectively.
    Considering the weighted average model parameters for the band structure, the energy gap is approximately $0.3$ meV, corresponding to a length scale of order $\lambda_F \sim 10~\mu$m. 
    Therefore, size effects suppress the boundary states features in the system considered here, where $d \ll \lambda_F$.

    \begin{figure}[tphb]
    	\center
    	\includegraphics[width=\linewidth]{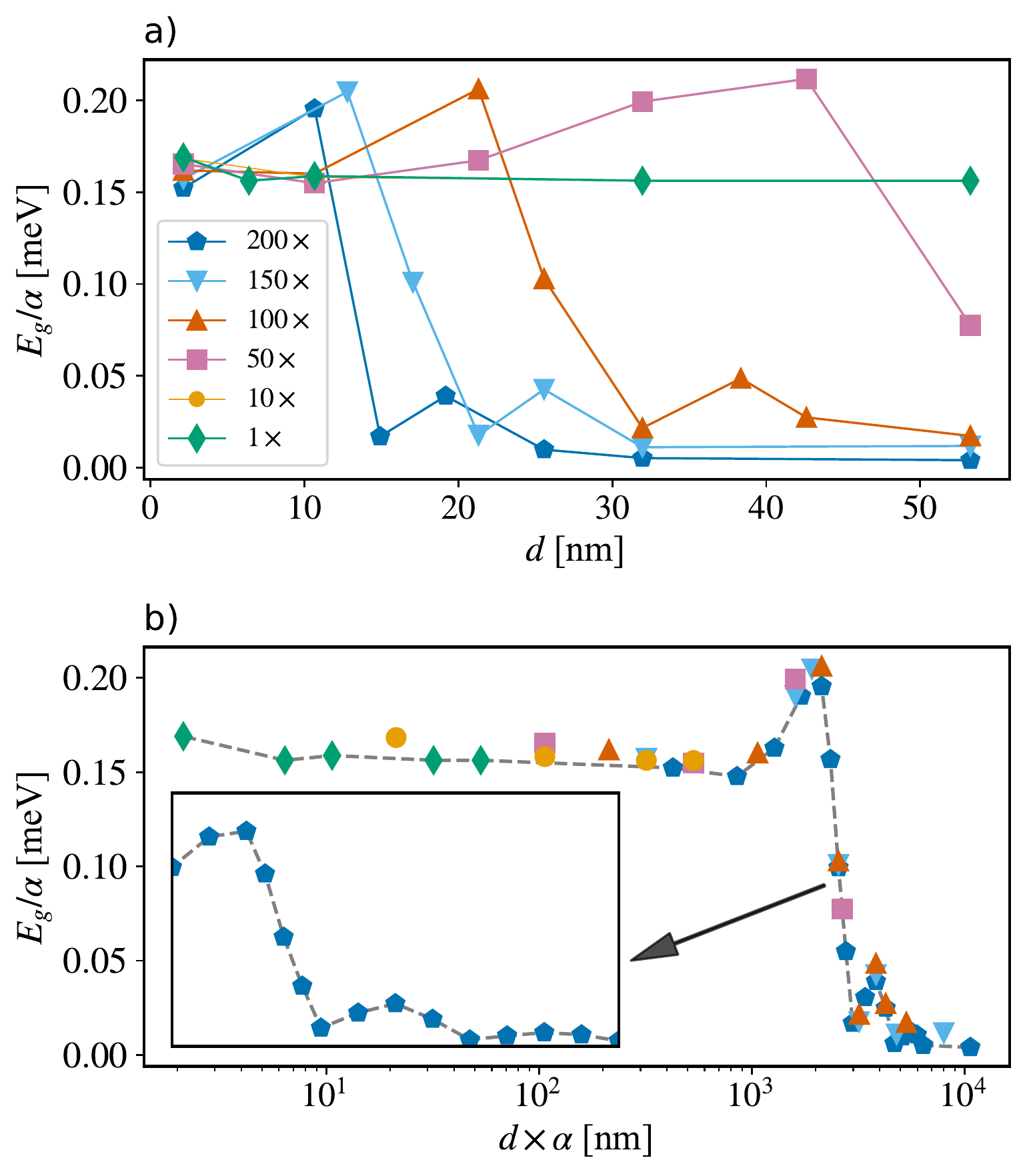}
    	\caption{\small (a) Energy gaps for Gr/(Mo--W)$_2$ as a function of domain size, normalised by the enhancement factor $\alpha=\{1,10,50,100,150,200\}$ of the proximity SOC parameters in the Gr/MoSe$_2$ and Gr/WSe$_2$ domains. (b) is the exact replica of (a) except that the domains are enhanced by $\alpha$ for better comparison between the different structures. The inset shows the zoomed-in oscillations of the energy gap for $\alpha = 200$.
    	\label{energy_gap}} 
    \end{figure}

    Our DFT results and simple scaling analysis suggest that interface states do not emerge at this system size, yet the valley Chern index analysis suggests that these states must eventually appear as the domain size is increased. 
    We now address the question of when in-gap states appear and how the band structure evolves between the homogeneous effective medium and multiple domain regimes.
    We start with small Gr/(Mo--W)Se$_2$ systems and gradually enlarge the domain size so that states in each domain are less likely to interact.
    We note that, since the TB band structure agrees exactly with the DFT results (See Fig.\ref{schematic} (b)), we can use the computationally less-expensive TB model discussed in Sec. \ref{tb_model} to evaluate the electronic structure of wider systems, where ab initio calculations are intractable.
    Fig. \ref{band_transform} presents the low energy dispersion of Gr/(Mo--W)Se$_2$ with different size domains (different columns).  
    The smallest cell case, shown in the leftmost panels, also plots the band structure predicted by the WAVG effective medium model (symbols) alongside the TB bands (solid lines) for comparison.
    From panel (a--d), we increase the domain size by a factor of 25, without any significant change to the TB band structures, which continue to exactly mimic the WAVG model.
    The emergent bands are not similar to those in pure MoSe$_2$ or WSe$_2$ domains (orange and cyan shaded regions in panel (d)) nor do they display any signatures of branch crossings which could indicate the emergence of localised interface modes \cite{touchais2022robust}. 
    Note that the domains here are still significantly smaller than the Fermi wavelength suggested by the energy gap scale in these systems. 
    To examine regimes where the Fermi wavelength and system size become similar in magnitude, we now increase the magnitude of each of the proximity terms in the Hamiltonian by a factor $\alpha = 50$ (middle row) and $\alpha=200$ (bottom row). 
    This approach essentially renormalises the energy and length scales and allows us to infer the physics of larger systems with realistic SOC parameters from that of smaller systems with enhanced SOC.
    For simplicity, we will continue to use the terminology Gr/MoSe$_2$ and Gr/WSe$_2$ to refer to the associated pristine systems with enhanced SOC, as the qualitative nature of the bands is unchanged.
    The validity of this approach will be discussed further in the next section.
    The smallest system cell for both $\alpha$ values, Fig. \ref{band_transform}(e) and (i), still perfectly follows the effective model band. 
    However, as we move towards larger domain sizes with enhanced SOC, the bulk gap changes and the bands evolve to form the in-gap states.
    As we increase the domain size for $\alpha=200$ in panels (i)--(l) , we note an initial enhancement of the energy gap, before eventually the gap closes due to branch crossings in the energy dispersion.
    The four branches crossing at zero-energy allude to the formation of interface modes, while the bulk bands closely follow those of pristine Gr/MoSe$_2$-type and Gr/WSe$_2$-type systems, as depicted by the shaded orange and cyan regions respectively. 
    The subgap states we find here are similar to those of a bilayer graphene with stacking registry domains in presence of an external electric field \cite{vaezi2013topological,yin2016direct}, and also the domain wall modes in proximitised graphene with a sign-changing VZ term on a constant Rashba background \cite{touchais2022robust}. 
    The initial enhancement of the gap, visible for example in Fig. \ref{band_transform}(j), is consistent with the fact that the gap for uniform Gr/MoSe$_2$-type and Gr/WSe$_2$-type systems is greater than that of the alloyed system.
    It can therefore be seen as a sign of the breakdown of alloyed-type behaviour and the onset of individual domains with different electronic properties.

    \begin{figure*}[tphb]
    	\center
    	\includegraphics[width=\linewidth]{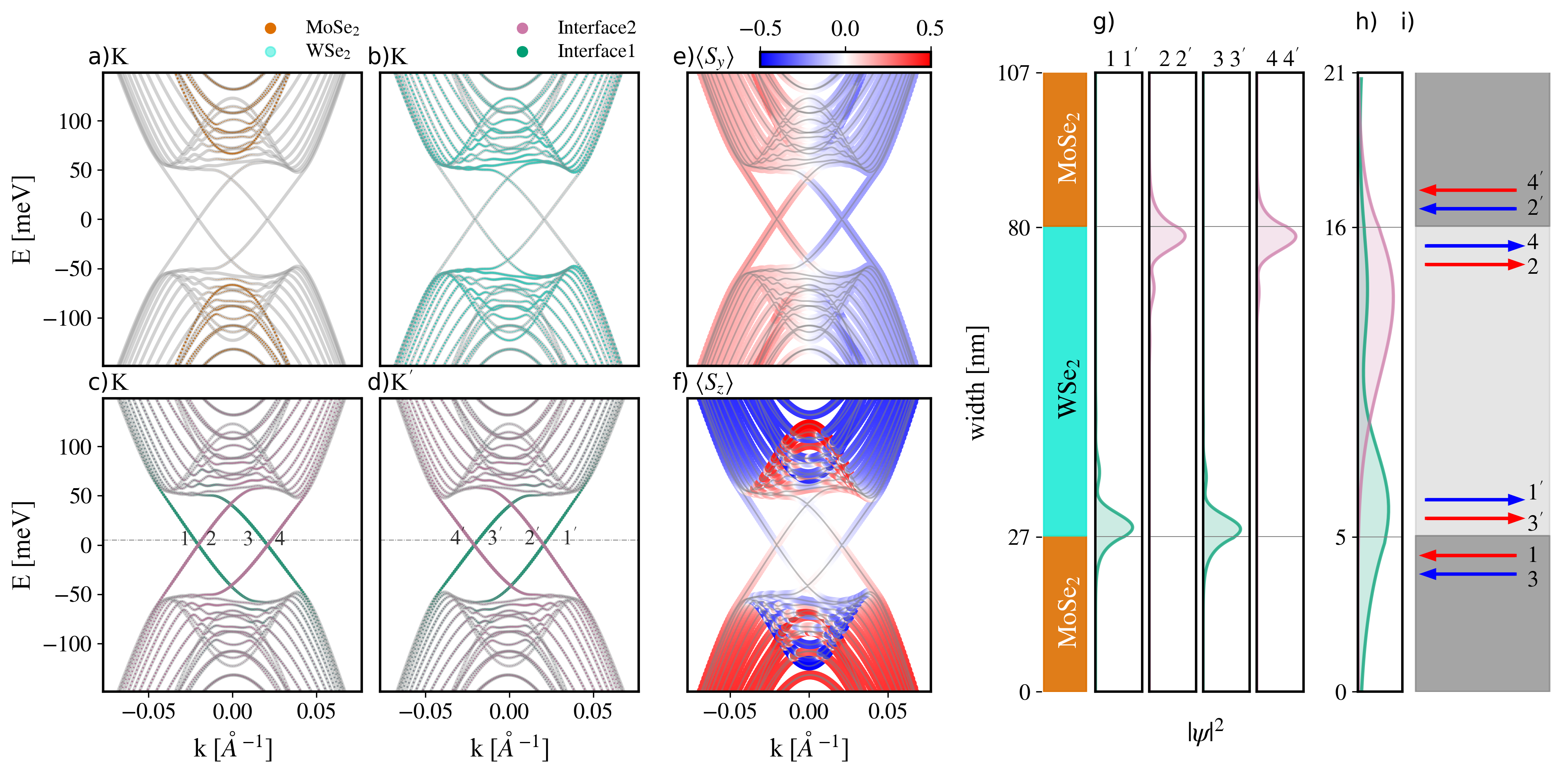}
    	\caption{\small Different domains, i.e. (a) Gr/MoSe$_2$, (b) Gr/WSe$_2$, and interfaces contribution to eigenvalue weight in a proximitised graphene with $\approx 107$ nm width around the (c) K and (d) K$^\prime$ valleys. The SOC parameters of MoSe$_2$ and WSe$_2$ are enhanced by a factor of 200. The contribution of the first and second interfaces is depicted in green and purple colors. (g) The real space expansion of the interface states at momentum values where the dashed line in (c,d) crosses the lowest conduction band. The colors correspond to the interface at which the in-gap states are localised. 
        (e,f) The spin textures along the in-plane (y) and transverse (z) axis. The band colors indicate the spin projection.
        (h) Real state distribution of the wave functions for a 21 nm cell system at k values indicated by empty red disks in fig. \ref{band_transform} (j). The solid lines in (g,h) depict the interfaces' location in the cell. (i) Schematic illustration of in-gap states propagating along the Mo/W interfaces at the K and K$^\prime$ valleys. The labels correspond to the energy levels depicted in panels (c,d). The in-gap states colors show the spin projection along the $y$ axis.  
    	\label{wf}} 
    \end{figure*}

    \section{Size effects}\label{size_eff}
    Fig. \ref{band_transform} suggests that proximitised graphene can indeed host branch crossings, and the associated interface states, at the interface between regions with MoSe$_2$-type and WSe$_2$-type proximity SOC. 
    However, this was only shown explicitly for systems with a hugely enhanced SOC, $\alpha=200$, so it is worth examining the agreement between systems with different lengths and energy scales more closely.
    In particular, it is important to determine the domain size that would be required for interface states to emerge in a real Gr/(Mo--W)Se$_2$ heterostructure, i.e. $\alpha=1$.
    
    In the simplest Dirac model for graphene, the inverse scaling relationship between length ($L$) and energy ($E$) is perfect: the mathematical description is identical if $E \rightarrow {\alpha E}$ and $L \rightarrow \frac{L}{\alpha}$.
    However, this is not guaranteed for more complicated dispersion relations or in TB descriptions, where the scale of the SOC parameters compared to the hopping integral also plays a role.
    It is therefore important to establish that similar behaviour is found across the different energy and length scales considered in Fig. \ref{band_transform}.
    One way to tackle this issue is to compare the band gap evolution as a function of both the domain size and the magnitude of the SOC terms. 
    Band gap closure is a clear fingerprint of the establishment of branch crossings in the electronic structure, as shown in Fig. \ref{band_transform}, so the band gap evolution is a suitable metric to ensure the same behaviour is expected across different scales.

    Fig. \ref{energy_gap}(a) shows the energy gaps for Gr/(Mo--W)Se$_2$ as a function of domain size, normalised by the enhancement factor $\alpha$ of the proximity SOC parameters in the Gr/MoSe$_2$ and Gr/WSe$_2$ regions. 
    Alongside the values in Fig. \ref{band_transform}, we also consider a range of intermediate values and show here results for $\alpha=\{1,10,50,100,150,200\}$. 
    The convergence of the curves for different SOC strengths at small $d$ shows that random alloy behaviour emerges for all cases, as expected, in the limit of very small domains.
    This effective medium regime also persists across all the domain sizes considered for small values of $\alpha$, as evident from the constant band gaps for the $\alpha=1$ and $\alpha=10$ curves.  
    However, for larger values of $\alpha$, a non-monotonic evolution of the band gap is observed.
    The dark blue $\alpha=200$ curve corresponds to the bottom row of Fig. \ref{band_transform}, and shows the initial increase and eventual suppression of the band gap as the domain size is increased.
    Similar behaviour is noted for the $\alpha=50, 100, 150$ curves, but with a shift towards larger domain sizes for smaller SOC parameters, as expected.

    To determine whether the band gap evolution is universal across different scales, in Fig. \ref{energy_gap}(b), we plot the band gaps found for a wide range of $d$ combinations.
    In this plot, we also re-scale the domain size according to the value of $\alpha$ used in the calculation and find that all the different cases in Fig. \ref{energy_gap}(b) now lie on a single `master' curve (dashed line). 
    This confirms that the simple inverse scaling relationship between energy and size holds for proximitised graphene. 
    We note this can be explained by the fact that although the SOC parameters are enhanced by up to two orders of magnitude, they are still significantly smaller than the magnitude of the hopping integral ($\sim 3$ eV) in graphene and the resultant gaps are formed well within the linear regime of the graphene dispersion relation.  
    The robustness  of this scaling also allows us to predict the domain size required for band gap closure and the emergence of branch crossings in real Gr/(Mo--W)Se$_2$ heterostructures.
    The smallest domain size required in this case is approximately $4\mu$m, which we note corresponds closely to the domain sizes achieved in CVD-grown MoSe$_2$/WSe$_2$ lateral heterostructures \cite{sahoo2018one}.

    Finally, we note that the band gap closure at larger system sizes does not occur as a simple decay, but instead a superimposed oscillation occurs which is shown more clearly in the inset of Fig. \ref{energy_gap}(b) for the $\alpha=200$ case. 
    This is consistent with a long-ranged coupling between localised states at neighbouring interfaces, similar to the Ruderman-Kittel-Kasuya-Yosida (RKKY) interaction that can occur between magnetic impurities \cite{vozmediano2005local, saremi2007rkky, power2013indirect} or defect lines \cite{gorman2014rkkylines} in graphene. 
    A decay rate of $\sim d^{-2.7}$ is extracted from the band gaps shown here, which is consistent with the expected decays of between $d^{-2}$ and  $d^{-3}$ predicted for different RKKY scenarios in graphene.

    \section{Interface states}\label{DW_modes}

    Proximity to either MoSe$_2$ and WSe$_2$ breaks inversion symmetry and leads to sharp peaks in the momentum-space Berry curvatures around the K and K$^\prime$ valleys in the energy dispersion of graphene.
    The proximitised Hamiltonian thus gives rise to nonzero Chern indices in the individual valleys.
    However, as a result of the unbroken time-reversal symmetry, these valley Chern indices are not independent due to the vanishing total Chern number and cannot be regarded as a robust topological order \cite{Alsharari2018Topological}.
    Nevertheless, at the interfaces of domains with different valley Chern indices, robust one-dimensional states form, however, they are not topologically protected against any perturbation that couples the valleys, i.e. intervalley scattering processes.
    In the absence of such scattering, the number of 1D channels per valley at such an interface is equal to the difference between their valley Chern indices, i.e. $C_V = C_K - C_{K^\prime}$.
    Since $C_V=-1$ for Gr/MoSe$_2$ and $C_V=1$ for Gr/WSe$_2$, we expect two pairs of boundary states for each interface in the system. 
    
    Since our supercell has two interfaces, we therefore expect four branch crossings in total per valley, consistent with what is observed in, for example, Fig. \ref{band_transform} (l).
    We now need to confirm that band structure indeed corresponds to a superposition of Gr/MoSe$_2$- and Gr/WeSe$_2$-type gapped bulk bands in their respective domains and gapless modes along the interfaces between them. 
    This can be seen by analysing the eigenvector weights in different regions of the supercell.
    The contributions of the MoSe$_2$ (orange), WSe$_2$ (cyan), first interface (interface1, green) and second interface (interface2, purple) to the heterostructure dispersion near the K point are shown by the coloured symbols in Fig. \ref{wf} (a--c).
    The bulk bands can clearly be seen to be a combination of independent (a) Gr/MoSe$_2$- and (b) Gr/WSe$_2$-type bands, each largely confined to its own domain. 
    For example, Fig. \ref{wf} (a) shows that the outermost, inverted subbands, which are a defining feature of the pure Gr/WSe$_2$ system, have no weight in Gr/MoSe$_2$ regions.
    Meanwhile, the low-energy branch crossings, numbered 1--4 in Fig. \ref{wf}(c), are strongly localised along the two interfaces in the supercell.
    Furthermore, we see the expected number of crossings at each interface with the propagation direction of interface modes swapped in the K$^\prime$ valley, Fig. \ref{wf} (d) due to time-reversal symmetry. 

    The spatial localisation of the interface branch modes is shown in more detail in Fig. \ref{wf} (g), where we plot the eigenvector weight as a function of position for each of the branch crossings numbered in Fig. \ref{wf} (c) and (d).
    The modes present at each interface, together with their propagation direction and spin orientation, are shown in Fig. \ref{wf} (i). 
    In each valley, there are two pairs of interface branches, which are localised at opposite interfaces and propagate in opposite directions.
    In the K valley, for example, modes 1 and 3 propagate from right to left along interface1, whereas modes 2 and 4 propagate in the opposite direction along interface2. 
    Fig. \ref{wf}(e) and (f) show the spin textures of the low-energy states along the in-plane ($y$) and out-of-plane ($z$) directions respectively.
    Unlike the edge modes in the simplest Quantum Spin Hall effect case \cite{kane2005quantum}, the low-energy interface modes here have largely in-plane spin polarisations, with only a smaller component in the $z$ direction due to the presence of the Rashba term in the proximity SOC. We have coloured the arrows in Fig. \ref{wf} (i) according to the larger, in-plane polarisation of each mode.
    Note that it is the valley and not the spin index which determines the propagation direction of boundary states. 
    The interfaces in this system can be viewed as hosting two instances of the quantum Valley Hall effect (VHE), one for each of two spin orientations.
    As such, transport in the interface modes is protected against backscattering in the absence of intervalley scattering. 
  
    For smaller domains, the wavefunctions of counter-propagating modes from neighbouring interfaces can overlap, as shown in Fig.~\ref{wf} (h) for the $d=10.7$ nm case.
    Here, the low energy states are delocalised over the entire graphene lattice and do not give rise to well-defined interface states. 
    The resultant hybridisation opens a gap in the corresponding band structure, which is shown in Fig.~\ref{band_transform}(j). 
    On the other hand, when the domains become large enough to prevent overlap, as shown previously in Fig. ~\ref{wf} (g), the interface modes emerge and the gap in the band structure closes.

    Finally, we note that in systems where the interfaces are oriented along the armchair direction, the K and K$^\prime$ valleys are folded onto the same point in the reciprocal space.
    However, similar to previous studies \cite{touchais2022robust,vaezi2013topological,jung2011valley}, our calculations show that this does not lead to the annihilation of the interface states. 
    Only in presence of a sharp transition between the different VHE regions, does a small bulk gap open but as the transition becomes smoother the gap vanishes.
   As a result, in the absence of intervalley scattering, the interface states protected by time-reversal symmetry are particularly resilient to orientational faults. This is especially important because it may be difficult to achieve precise alignment between the TMDC interface and the graphene zigzag direction in heterostructures fabricated using dry transfer methods.

    \section{Conclusions}\label{conclusions}
    Proximity to TMDC can lead to the emergence of TPH states in graphene similar to those predicted by Kane and Mele when spin-orbit interactions are included. 
    These interactions are rather complex and consist of different induced coupling mechanisms in the low-energy spectrum. 
    The SOC in proximitised graphene involves substrate-induced Rashba term that couples electrons of different spin orientations, the intrinsic term that induces a topological gap, and the sublattice-asymmetric VZ term that couples the spin and valleys. In contrast to Kane-Mele model, however, the staggered intrinsic SOC in Gr/TMDC yields a trivial topological order, with nonzero Berry curvature and hence Chern index at individual valleys.
    In this work, we have studied the case of Gr/MoSe$_2$, with a direct optical band gap and Gr/WSe$_2$ with VZ-driven inverted bands since they can act as an interesting platform for the realization of TPH states by enabling a topological crossover with band gap closure when placed together in a composite system.
    We demonstrated the emergence of the localised boundary states in a Gr/(Mo--W)Se$_2$ heterostructure where W- and Mo-like domains with opposite valley Chern indices are formed. Utilising a microscopic study we showed how topologically protected in-gap modes evolve with the lattice size, and how they can hybridize to a weighted average effective model state similar to a randomized composite system when the domains are remarkably smaller than the Fermi wavelength.

    \begin{acknowledgments}
		The authors acknowledge the support of the Irish Centre for High-End Computing (ICHEC). This study is funded by the Irish Research Council under the Laureate awards and the Government of Ireland postdoctoral fellowship program.  
	\end{acknowledgments}

    \section{References}

	%

\end{document}